\documentclass[runningheads]{llncs}
\usepackage{graphicx}

\usepackage[linesnumbered,ruled,vlined]{algorithm2e}
\usepackage{amsmath, amssymb, amsfonts}
\usepackage{subfig}
\usepackage{listings}
\usepackage{mathtools}
\newtheorem{observation}{Observation}
\usepackage{cite}

\usepackage{comment}
\usepackage{graphicx}
\begin{document}
\title{Towards Stronger Blockchains: Security Against Front-Running Attacks}
%
%
\author{Anshuman Misra \and
Ajay D. Kshemkalyani
}
\titlerunning{Towards Stronger Blockchains: Security Against Front-Running Attacks}
\authorrunning{A Misra and AD Kshemkalyani}
%
\institute{University of Illinois at Chicago \\
\email{amisra7@uic.edu},
\email{ajay@uic.edu}}
\maketitle              
\begin{abstract}

Blockchains add transactions to a distributed shared ledger by arriving at consensus on sets of transactions contained in blocks. This provides a total ordering on a set of global transactions. However, total ordering is not enough to satisfy application semantics under the Byzantine fault model. This is due to the fact that malicious miners and clients can collaborate to add their own transactions ahead of correct clients' transactions in order to gain application level and financial advantages. These attacks fall under the umbrella of front-running attacks. Therefore, total ordering is not strong enough to preserve application semantics. In this paper, we propose causality preserving total order as a solution to this problem. The resulting Blockchains will be stronger than traditional consensus based blockchains and will provide enhanced security ensuring correct application semantics in a Byzantine setting.

\end{abstract}

\begin{section}{Introduction}

Blockchain is a shared distributed ledger that provides a tamper-proof ordered sequence of records. Bitcoin \cite{nakamoto2008bitcoin} was the first blockchain that provided a solution to the \textit{double-spending problem} and revolutionized electronic money transfer. Bitcoin solves Byzantine-tolerant consensus \cite{lamport2019byzantine} via proof-of-work. This led to the development of further blockchains that solved consensus such as Ethereum \cite{wood2014ethereum} that go a step further and provide smart contracts \cite{dickerson2017adding} that allow the blockchain to act as a \textit{universal computer}. Smart contracts are code hosted on the blockchain that provide operations to change the state of the blockchain. Since the code is tamper-proof, a set of parties can conduct business in a transparent manner on the blockchain. Further, smart contracts provide the capability to run classic centralized applications on the blockchain in a decentralized manner, such as auctions \cite{hahn2017smart} and elections \cite{hanifatunnisa2017blockchain}. As more applications are designed for blockchain, an important question that arises is --- does blockchain provide the required semantics for these applications? In the case of peer-to-peer money transfer, the answer is yes, because total order prevents double-spending. However, total order is not enough for a decentralized auction, because a Byzantine miner can collude with a Byzantine client by informing the client of its opponents' bids prior to them being added to the blockchain. This lack of enforcement of semantics provides an opportunity to Byzantine nodes to launch a variety of attacks known as front-running attacks \cite{eskandari2020sok,zust2021analyzing}. In this paper, we propose utilizing causal ordering protocols enforcing strong safety \cite{misra2022solvability,misra2023byzantine,misra2023byzantine2} to provide an enhanced level of security in the blockchain ecosystem. Our contributions are as follows:

\begin{enumerate}
    \item We formalize front-running attacks and prove that they are a violation of causal ordering.

    \medskip

    \item We prove that utilizing a causal ordering protocol will enforce application semantics and make the blockchain more secure and suitable for classic centralized applications.

    \medskip

    \item We introduce a protocol to provide security against front-running attacks by providing a causality preserving total ordering across transactions recorded in the blockchain. We term the resulting blockchain as a strong blockchain since it provides stronger security guarantees and semantics compared to traditional blockchains.

    \medskip

    \item We prove the correctness of our protocol and analyze its intrinsic fairness properties.

\end{enumerate}

\end{section}

\begin{section}{System Model}
\label{sec:systemmodel}

This paper models the set of miners as a distributed system having Byzantine processes which are processes that can misbehave \cite{DBLP:journals/jacm/PeaseSL80,DBLP:journals/toplas/LamportSP82}. A correct process (miner) behaves exactly as specified by the blockchain protocol whereas a Byzantine process (miner) may exhibit arbitrary behaviour including crashing at any point during the execution. A Byzantine process cannot impersonate another process or spawn new processes. The distributed system is modelled as an undirected graph $G = (\mathcal{M},H)$. Here $\mathcal{M}$ is the set of miners adding blocks to the blockchain.  Let $n$ be $|\mathcal{M}|$. $H$ is the set of FIFO logical communication links over which miners communicate by message passing. $G$ is a complete graph. This model is equivalent to the permissioned blockchain model \cite{polge2021permissioned}. Nonetheless, as can be seen by the proofs in this paper, our results hold for permissionless blockchains \cite{nakamoto2008bitcoin} as well. However, the solution we provide is geared towards permissioned blockchains.

The system is assumed to be synchronous, i.e., there is a known fixed upper bound $\delta$ on the message latency, and a known fixed upper bound $\psi$ on the relative speeds of processors \cite{DBLP:journals/jacm/DworkLS88}. This is opposed to an asynchronous system, i.e., there is no upper bound $\delta$ on the message latency, nor any upper bound $\psi$ on the relative speeds of processors \cite{DBLP:journals/jacm/DworkLS88}. Clients send their transactions to the system of miners by broadcasting protocol message containing the transaction to the system. This message contains all the required metadata for the transaction such as gas fees and the client's identity. Next,
transactions sit at each miner's mempool \cite{saad2019exploring}, waiting to get added to the blockchain. Clients can also be Byzantine and collude with Byzantine miners and miners can also act as clients in the system. Our protocol assumes an upper bound on the number of Byzantine miners, $t$ with $n \geq 3t + 2$. The number of Byzantine clients is assumed to be unbounded.

\begin{definition} 
\label{definition:hb}
The happens before relation $\rightarrow$ on messages consists of the following rules:
\begin{enumerate}
\item The set of messages delivered from any process $p_i$  by a process is totally ordered by $\rightarrow$.
\item If $p_i$ sent or delivered message $m$ before sending message $m'$, then $m \rightarrow m'$.
\item If $m\rightarrow m'$ and $m'\rightarrow m''$, then $m\rightarrow m''$.
\end{enumerate}
\end{definition}

Let $R$ denote the set of messages in the execution.

\begin{definition}
The \textit{causal past} of message $m$ is denoted as $CP(m)$ and defined as the set of messages in $R$ that causally precede message $m$ under $\rightarrow$.
\end{definition}

The correctness of Byzantine causal order unicast/multicast/broadcast 
is specified on $(R,\rightarrow)$ for strong safety.

\begin{definition}
\label{definition:ss+l}
A causal ordering algorithm for unicast/multicast/broadcast messages must ensure the following:
\begin{enumerate}
    \item \textbf{Strong Safety:} $\forall m' \in CP(m)$ such that $m'$ and $m$ are sent to the same (correct) process(es), no correct process delivers $m$ before $m'$.
    \item \textbf{Liveness:} Each message sent by a correct process to another correct process will be eventually delivered.
\end{enumerate}
\end{definition}

\begin{definition}

\label{def:transaction}

A \textbf{transaction} is a string contained in messages broadcasted to the system of miners with the intention of being recorded in the blockchain via consensus. Transactions change the state of the blockchain by executing business logic between two or more clients.

\end{definition}

\begin{definition}
    \label{def:hb_tx}
    The happens before relation $\rightarrow$ on transactions is defined as follows: Given messages $m_1$ and $m_2$ containing transactions $t_1$ and $t_2$ respectively, $t_1 \rightarrow t_2$ if and only if $m_1 \rightarrow m_2$. 
    
\end{definition}

\begin{definition}
The \textit{causal past} of transaction $t$ is denoted as $CP(t)$ and defined as the set of transactions that causally precede $t$ under $\rightarrow$.
\end{definition}

\begin{definition}
\label{definition:ss+l_tx}
A causal ordering algorithm for blockchain $BT$ must ensure the following:
\begin{enumerate}
    \item \textbf{Strong Safety:} Given transaction $t$, $\forall t' \in CP(t)$, $t'$ gets recorded in $BT$ before $t$.
    \item \textbf{Liveness:} Each transaction sent by a correct client eventually arrives in every correct miner's memory pool.
\end{enumerate}
\end{definition}

\begin{definition}

\label{def:block}

A \textbf{block} contains a sequence of totally ordered transactions and a hash of its parent block. A block is only added to the blockchain after the system of miners arrive at consensus on the contents of the block.

\end{definition}

\begin{definition}

\label{def:blockchain}

\textbf{Blockchain} is a distributed data structure consisting of a tree of blocks. Each block has only one parent (except the genesis block) and may have multiple children blocks.

\end{definition}

\begin{definition}

\label{def:blockchain}

Given a tree $BT$ containing a blockchain, the \textbf{consensus chain} is a ordered sequence of blocks $B_0, B_1, ...., B_l$ such that $B_k$ is the parent of $B_{k+1}$ and $tree\_depth(BT) = n$.

\end{definition}

\begin{definition}

\label{def:valid_blockchain}

$BT$ is a \textbf{valid blockchain} if $BT$ contains one and only one consensus chain.

\end{definition}

\end{section}

\begin{section}{Front-Running Attacks}

In this section we first present a broad family of attacks called front-running attacks. We formalize the attack and prove that it is essentially an attack on causal ordering. An important point to note is that front-running attacks are executed prior to execution of the blockchain consensus protocol. Miners can view unconfirmed transactions in their memory pools and broadcast their own transactions with higher transaction fees with the intention of executing front-running attacks on unsuspecting clients. Byzantine miners can also collude with Byzantine clients to execute front-running attacks. However, without loss of generality in our proofs and solutions, we assume that miners act as clients in executing attacks. The following are illustrative examples of front-running attacks on real-world applications:

\begin{enumerate}
    \item An honest client process $p_i$ sends transaction $t_i$ to the network as part of a decentralized auction. A malicious miner $M$ reads $t_i$, figures out the bid value $p_i$ wants to place for an asset being auctioned and sends its own transaction $t_M$ with the purpose of getting into the blockchain first. $t_M$ would contain a higher bid price than $p_i$'s bid price. This results in an unfair advantage for $M$ in winning the auction.

    \medskip

    \item An honest client process $p_i$ sends a request to buy cryptocurrency (transaction $t_1$) at price $x$, where the market price $y$ is less than $x$. A malicious miner $M$ can aim to profit off this by adding two transactions to the block where it includes $t_1$:

    \begin{enumerate}
        \item It adds $t_0$ buying cryptocurrency at price $y$ from the market. $t_0$ is placed before $t_1$.

        \item It sells cryptocurrency to $p_i$ in transaction $t_2$ to $p_i$ (placed after $t_1$) at price $x$.
    \end{enumerate}

    This results in $M$ making a profit by arbitraging off an honest client with a profit of $(y-x)$ per coin.
    
\end{enumerate}

An honest miner should not look into the content of transactions in the network. Blockchains incentivize miners to go after transactions with higher mining fees to maximize profits. The blockchain protocol requires miners to be concerned with only transaction fees and not the contents of transactions. The Byzantine fault model encapsulates behaviour that does not follow specified protocol. Therefore, such malicious miners can be modeled as Byzantine processes and Byzantine fault-tolerant protocols can be utilized to prevent such behaviour. 

\begin{observation}
    \label{o:byz}
    Miners executing front-running attacks are Byzantine.
\end{observation}

Front-running attacks are broadly categorized as follows \cite{torres2021frontrunner}:

\begin{enumerate}
    \item \textbf{Displacement Attack}: A Byzantine miner reads transaction $t$ from its memory pool and broadcasts its own transaction (copying contents of $t$) $t'$ with higher transaction fees in order to record $t'$ in the blockchain before $t$. 
    
    \item \textbf{Sandwich Attack}: A Byzantine miner reads transaction $t_1$ from its memory pool and broadcasts two transactions $t_0$ and $t_2$ with the intention of recording $t_0$ before $t_1$ and $t_2$ after $t_1$ in the same block. In this way, the Byzantine miner creates an arbitrage opportunity to make a profit.
    
    \item \textbf{Suppression Attack}: A Byzantine miner reads transaction $t$ from its memory pool and broadcasts a set of transactions $T$ containing transactions with  high transaction fees. This attack essentially forces $t$ to not get recorded in the next block in the blockchain. 

\end{enumerate}

We now formally define front-running attacks and prove that they require causal ordering violation in order to occur. Since front-running attacks are executed before consensus and are harder to execute when no forks exist, for the sake of proofs we assume without loss of generality that they are executed on valid blockchains. 

\begin{definition}
    \label{def:front_run}
    
    A Byzantine miner executes a \textbf{front-running attack} by reading an unconfirmed transaction $t_x$ and broadcasting/mining its own transaction $t_y$ with the intention of recording $t_y$ before $t_x$ in the consensus chain of a valid blockchain $BT$.
    
\end{definition}

\begin{theorem}
    \label{th:front_run}

    Front-running attacks are a violation of causal ordering.
    
\end{theorem}

\begin{proof}

$BT$ is a valid blockchain and $C = consensus\_chain(BT)$.
Let $p_i$ broadcast $m_1$ (containing transaction $t_1$) to the system of miners. Miner $M$ delivers $m_1$ and adds $t_1$ to its memory pool. Miner $M$ then broadcasts $m_2$ (containing $t_2$) with significantly higher transaction fees than $m_1$ ($t_1$), with the intention of adding $t_2$ to $C$ before $t_1$ is added to it. If this attack succeeds, one of the following scenarios may play out:

\begin{enumerate}
    \item Miner $M'$ ($M$ may be $M'$) succeeds in adding the next block $B$ containing $t_2$ to $C$. The new consensus chain of $BT$ is $C' = C + B$. Eventually, $t_1$ gets added to $consensus\_chain(BT)$ as part of block $B'$. Since $C'$ is a prefix of some future consensus chain, $consensus\_chain(BT)$, $t_2$ is ordered before $t_1$.

    \item Miner $M'$ ($M$ may be $M'$) succeeds in adding the next block $B$ to $C$. $B$ contains both $t_1$ and $t_2$ with $t_2$ ordered before $t_1$.
    
\end{enumerate}

By Definition \ref{def:front_run}, this is a front-running attack on $t_1$ by $M$ via $t_2$. In order to execute this attack, $M$ delivered $m_1$ and broadcasted $m_2$. By the message order rule in Definition \ref{definition:hb}, $m_1 \rightarrow m_2$. Since $t_2$ is recorded in $C$ before $t_1$, the contents of $m_2$ are consumed by the system before the contents of $m_1$ resulting in a strong safety violation as per Definition \ref{definition:ss+l}. Therefore, it is clear that a front-running attack across transactions requires a causality violation across their respective protocol messages. $\square$

\end{proof}

\end{section}

\begin{section}{Background}
\label{sec:background}

\subsection{Some Cryptographic Basics}
\label{sec:crypto}

We utilize non-interactive threshold cryptography as a means to guarantee strong safety of multicasts \cite{shoup2002securing}. Threshold cryptography consists of an initialization function to generate keys, message encryption, sharing decrypted shares of the message and finally combining the decrypted shares to obtain the original message from ciphertext. The following functions are used in a threshold cryptographic scheme:

\begin{definition}
    \label{def:initialize}
    The dealer executes the generate() function to obtain the public key $PK$, verification key $VK$ and the private keys
    $SK_0$, $SK_1$, ... , $SK_n$.

\end{definition}

The dealer shares private key $SK_i$ with each process $p_i$ while $PK$ and $VK$ are publicly available.

\begin{definition}
    \label{def:generate}
    When process $p_i$ wants to send a message $m$ to $p_j$, it executes $E(PK,m,L)$ to obtain $C_m$. Here $C_m$ is the ciphertext corresponding to $m$, $E$ is the encryption algorithm and $L$ is a label to identify $m$. $p_i$ then broadcasts $C_m$ to the system of processes.
\end{definition}

\begin{definition}
    \label{def:share}
    When process $p_l$ receives ciphertext $C_m$, it executes $D(SK_l,C_m)$ to obtain $\sigma^m_l$ where $D$ is the decryption share generation algorithm and $\sigma^m_l$ is $p_l$'s decryption share for message $m$.
\end{definition}

When process $p_j$ receives a cipher message $C_m$ intended for it, it has to wait for $k$ decryption shares to arrive from the system to obtain $m$. The value of $k$ depends on the security properties of the system. It derives the message from the ciphertext as follows:

\begin{definition}
    \label{def:combine}
    When process $p_j$ wants to generate the original message $m$ from ciphertext $C_m$, it executes $C(VK,C_m,S)$ where S is a set of $k$ decryption shares for $m$ and $C$ is the combining algorithm for the $k$ decryption shares. 
\end{definition}

The following function is used to verify the authenticity of a decryption share:

\begin{definition}
    \label{def:verify}
    When a decryption share $\sigma$ is received for message $m$, the Share Verification Algorithm is used to ascertain whether $\sigma$ is authentic : $V(VK,C_m,\sigma) = 1$ if $\sigma$ is authentic, $V(VK,C_m,\sigma) = 0$ if $\sigma$ is not authentic.
\end{definition}

\subsection{Byzantine Causal Broadcast via Byzantine Reliable Broadcast}
\label{sec:btmb}

We propose a causal order broadcast algorithm for clients to send transactions to miners. Byzantine-tolerant causal broadcast is invoked as {\sf BC\_broadcast($m$)} and delivers a message through {\sf BC\_deliver($m$)}. 

\begin{definition}
\label{definition:BCM0}
Byzantine Causal Multicast satisfies the following properties: 
\begin{enumerate}
\item (BCB-Validity:) If a correct process $p_i$ {\sf BC\_deliver}s message $m$ from $send\-er(m)$ then $sender(m)$ must have {\sf BC\_broadcast} $m$.

\item (BCB-Termination-1:) If a correct process {\sf BC\_broadcast}s a message $m$ then some correct process eventually {\sf BC\_deliver}s $m$.

\item (BCB-Agreement or BCM-Termination-2:) If a correct process {\sf BC\_deliver}s a message $m$ from a possibly faulty process, then all correct processes eventually deliver $m$.

\item (BCB-Integrity:) For any message $m$,
every correct process $p_i$ {\sf BC\_deliver}s $m$ at most once. 

\item (BCB-Causal-Order:) If $m \rightarrow m'$, then no correct process {\sf BC\_deliver}s $m'$ before $m$.
\end{enumerate}
\end{definition}

BCB-Causal-Order is the Strong Safety property of Definition~\ref{definition:ss+l}. BCB-Termination-1 and BCB-Agreement imply the liveness property of Definitions~\ref{definition:ss+l}.

The Byzantine-tolerant Reliable Broadcast (BRB) \cite{bracha1987asynchronous,bracha1985asynchronous} is invoked by {\sf BR\_broadcast} and its message is delivered by {\sf BR\_deliver}, and satisfies the properties given below.

\begin{definition}
\label{definition:BRB0}
Byzantine-tolerant Reliable Broadcast (BRB) provides the following guarantees \cite{bracha1987asynchronous,bracha1985asynchronous}: 
\begin{enumerate}
\item (BRB-Validity:) If a correct process {\sf BR\_deliver}s a message $m$ from $sender\-(m)$, then $sender(m)$ must have {\sf BR\_broadcast} $m$.

\item (BRB-Termination-1:) If a correct process {\sf BR\_broadcast}s a message $m$, then some correct process eventually {\sf BR\_deliver}s $m$.

\item (BRB-Agreement or BRB-Termination-2:) If a correct process {\sf BR\_deliver}s a message $m$ from a possibly faulty process, then all correct processes eventually {\sf BR\_deliver} $m$.
\item (BRB-Integrity:) For any message $m$,
every correct process {\sf BR\_\-deliver}s $m$ at most once. 
\end{enumerate}
\end{definition}

\end{section}

\begin{section}{Causal Ordering Protocol to Prevent Front-Running Attacks}

\label{sec:multicastround}

In light of the result of Theorem \ref{th:front_run}, we present a causality preserving blockchain protocol to strengthen the security of blockchain to withstand front-running attacks under the synchronous system setting. A synchronous system can assume lock-step execution in rounds. Within a round, a process can send messages, then receive messages, and lastly have internal events; further a message sent in a round is received in the same round at all its destinations. Algorithm \ref{alg:sync_rounds} serves as a reference point for synchronous round-based communication. Without loss of generality, we assume that all processes send their messages at the beginning of each round, all messages arrive in the same round that they are sent out and messages are delivered at the end of each round. Threshold cryptography in conjunction with the execution in rounds and Byzantine Reliable Broadcast are used to ensure strong safety + liveness. Clients broadcast transactions to the system of miners encapsulated in protocol messages via BRB. Using BRB protects against liveness attacks by Byzantine clients via \textit{BRB-Termination-1} and \textit{BRB-Termination-2}.

\begin{algorithm}
\SetAlgoLined
{\small
\SetKwComment{Comment}{$\triangleright$ }{}
\KwData{Each process locally maintains two FIFO queues $Q_s$ and $Q_d$ for storing outgoing/incoming messages respectively}


\textbf{when} round $r$ starts: \\

\Indp

broadcast all messages in FIFO order after dequeuing from $Q_s$

\Indm

\textbf{when} round $r$ ends: \\

\Indp

deliver all messages in FIFO order after dequeuing from $Q_d$

\Indm

\textbf{when} the application is ready to broadcast message $m$: 

\Indp

$Q_s.enqueue(m)$

\Indm

\textbf{when} message $m$ arrives: \\
\Indp

$Q_d.enqueue(m)$

\Indm

\caption{Synchronous round-based message passing protocol}
\label{alg:sync_rounds}

}
\end{algorithm}

We present our solution, called \textit{causality preserving blockchain protocol} in Algorithm \ref{alg:co_protocol}. The blockchain consensus protocol generates a total ordering  of transactions. Our protocol guarantees a stronger property, \textit{causally consistent total ordering} of transactions. Algorithm \ref{alg:co_protocol} not only provides a total ordering of blockchain transactions, but also ensures that this total ordering does not violate causality across transactions, hence ensuring application semantics in a Byzantine setting. This is why we term any blockchain following our protocol as a stronger blockchain than classic blockchains. For simplicity, we term this as a \textit{strong blockchain} and is defined below.

\begin{definition}
    \label{def:strong_bc}
    A \textbf{strong blockchain} $BT$ must satisfy the following properties:

    \begin{enumerate}
        \item $BT$ is a valid blockchain

        \item $\forall t_1,t_2$ such that $t_1 \rightarrow t_2$, $t_1$ is recorded before $t_2$ in $BT$'s consensus chain.
    \end{enumerate}
    
\end{definition}

Algorithm \ref{alg:co_protocol} does not change the consensus mechanism and is therefore independent of the consensus protocol employed by the blockchain. Instead, Algorithm \ref{alg:co_protocol} ensures that only transactions whose causal past is already in the consensus chain are allowed to be mined. In other words, we have added a clause to determine whether a transaction is valid.

\begin{algorithm}[th!]
{\small 
\SetKwComment{Comment}{$\triangleright$ }{}
\KwData{Each client process $p_{c_i}$ has access to $PK$ (global public key), each miner process $p_m$ has access to $VK$ (global verification key). Each miner $p_{M_i}$ has access to a local secret key $SK_i$. Each client uses a FIFO queue $Q_s$ for outgoing protocol messages. Each miner $p_{M_i}$'s memory pool is denoted by a set $MP$. The causal past of transaction $t$ is denoted as $CP(t)$. The set of all miners is $\mathcal{M}$. $BT$ is the shared blockchain. $B^i_{r}$ is the block proposed by miner $p_{M_i}$ in round $r+1$.}
\medskip


\underline{\textbf{when} round $r$ starts at client $p_{c_i}$:} \\


\While{$Q_s.head() \neq \phi$}{

$C_m = Q_s.pop()$ \\
{\sf BR\_broadcast}$(C_m,M)$

}


\medskip

\underline{\textbf{when} client $p_{c_i}$ sends $m$ to $M$ via {\sf BC\_broadcast($m,M$)} in round $r$:} \\
$C_m = E(PK,m,id_m)$ \\
$Q_s.push(C_m)$


\medskip

\underline{\textbf{when} round $r$ starts at miner $p_{M_i}$:} \\

$B = consensus(candidate\_set)$ \Comment*[r]{consensus on the set of blocks delivered in the previous round}

$candidate\_set = \phi$ 

Add $B$ at the end of $consensus\_chain(BT)$

\For{all $t \in B$}{

delete $t$ from $MP$

\For{all $t'$ such that $t \in CP(t')$}{

$CP(t') = CP(t') \setminus t$

}

}

$B^i_r = \phi$

\For{all $t$ in $MP$ such that $t$ is semantically invalid}{
delete $t$

}

\For{all $t$ in $MP'$ where $MP' \subseteq MP$ $\land$ $CP(t) = \phi$}{

$B_r^i = B_r^i \cup \{t\}$ \Comment*[r]{Block construction with safe transactions}

}

\For{all $p_{m_j} \in \mathcal{M}$}{

send $B^i_{r}$ to $p_{m_j}$

}

\medskip

\underline{\textbf{when} $B_{r}^{j}$ arrives at miner $p_{M_i}$ during round $r$:} \Comment*[r]{Block created by miner $p_{m_j}$ in round $r$ and proposed for consensus in round $(r+1)$}

\For{all $t \in B_{r}^{j}$}{

\If{$t$ is semantically invalid $\lor$ $CP(t) \neq \phi$}{

discard $B_{r}^{j}$

}

}

\If{$B_{r}^{j}$ has not been discarded}{

$candidate\_set = candidate\_set \cup B_{r}^{j}$ \Comment*[r]{all safe blocks arriving in round r are added to candidate\_set}

}

\medskip

\underline{\textbf{when} $C_m$ is {\sf BR\_deliver}ed at miner $p_{M_i}$ in round $r$:}\\
$\sigma_i^m = D(SK_i,C_m)$ \\

\For{all  $p_{m_j} \in M$}{
   send $\sigma^m_i$ to $p_j$ in round $(r+1)$ 

}


\medskip

\underline{\textbf{when} miner $p_{M_i}$ receives $(2t+1)$th valid $\langle \sigma_x^m \rangle$ message by round $r$:} \\
Store $(t+1)$ decryption shares in set $S$ \\
$m = C(VK,C_m,S)$ \\
extract $t_m$ from $m$ \Comment*[r]{bc\_delivery(m)}
$CP(t_m) = MP$ \\
$MP = MP \cup \{t_m\}$ 

}
\caption{Causality Preserving Blockchain Protocol}
\label{alg:co_protocol}
\end{algorithm}

In addition to blockchain-specific properties that need to be satisfied (e.g., sufficient balance, identity of client), a transaction is not considered valid if it causally depends on one or more transactions that have not been finalized in the blockchain. We call valid transactions as safe transactions, formalized below:

\begin{definition}
    \label{def:valid_tx}
     Transaction $t$ is an \textbf{safe transaction} with regards to a valid blockchain $BT$ if and only if $\forall t' \in CP(t), \exists B \in  consensus\_chain(BT)$ such that $t' \in B$.

\end{definition}

Algorithm \ref{alg:co_protocol} only considers \textit{safe blocks} for consensus, preventing front-running attacks by maintaining causal relations across transactions. Definition \ref{def:valid_blk} formalizes a safe block. 

\begin{definition}
    \label{def:valid_blk}

    A \textbf{safe block} $B$ only contains valid transactions, or in case any invalid transaction $t \in B$, then $\forall t' \in CP(t)$, $t'$ must precede $t$ in $B$.

\end{definition}

Algorithm \ref{alg:co_protocol} provides the {\sf BC\_broadcast} primitive to clients to protect against front-running attacks and {\sf BC\_deliver} to miners for extracting transactions from messages. {\sf BR\_broadcast} and {\sf BR\_delivery} are the underlying primitives implementing Byzantine reliable broadcast (BRB) \cite{bracha1985asynchronous,bracha1987asynchronous}. Let $\beta$ and $\gamma$ denote the maximum and minimum number of rounds (sequential steps) respectively in a BRB protocol. For example, Bracha's BRB has $\beta=4$,$\gamma=3$ and requires $n> 3t$  \cite{bracha1985asynchronous,bracha1987asynchronous} whereas Imbs-Raynal \cite{imbs2016trading} has $\beta=3$,$\gamma=2$ and requires $n>5t$. A {\sf BR\_broadcast} sent in round $r$ is delivered as {\sf BR\_deliver} by round $r+\beta-1$. 
Although a message $m$ sent in a round is delivered after all messages sent in previous rounds, a Byzantine miner can peek into $m$ before its transaction is committed to the blockchain and send a causally dependent message $m'$ in the same round to initiate a broadcast send via its own {\sf BR\_broadcast}. $m'$ may be {\sf BR\_deliver}ed in the same round as $m$ at some miners, thus leading to a potential front-running attack across the transactions contained in $m$ and $m'$. To prevent a Byzantine process from peeking into the transaction of a message during the $\beta$ rounds, the message is encrypted using threshold encryption. In round $r+\beta$, each process that has {\sf BR\_deliver}ed the encrypted message sends its decryption share to the destinations of the multicast. A message gets revealed only at the end of round $r+\beta$.
Any message sent before that cannot be causally dependent on this revealed message $m$, and the only messages that the process sends that are causally dependent on the above message $m$ can get sent (and hence delivered) only in later rounds. This straightforwardly guarantees strong safety and liveness. The $\beta+1$ rounds $a\times (\beta+1) + 1$, $a\times (\beta+1) + 2$, $\ldots$ $a\times (\beta+1) + k$, $a\times (\beta+1) + \beta + 1$ constitute a meta-round $a$ for $a \geq 0$. Thus, rounds $r, r+1, \ldots r+\beta$, such that $r\, div\, (\beta+1) = a$ and $r \mod (\beta+1) = 0$ constitute meta-round $a$. The first $\beta$ rounds of a meta-round are for BRB and the $\beta+1$th round is for sending the decryption shares to the system of miners. 

Algorithm \ref{alg:co_protocol} consists of both miner side code and client side code divided into six \textit{when blocks}, each in reaction to an event in the protocol. The \textit{when block} from lines 1-4 is executed in the beginning of a round, with each client broadcasting messages using BRB it created in the previous round in a FIFO manner from a local queue containing those messages. FIFO ordering at the client in conjunction with FIFO channels ensures source order at the miners' end. The \textit{when block} in lines 5-7 describes how clients utilize the {\sf BC\_broadcast} primitive provided by Algorithm \ref{alg:co_protocol}. Clients encrypt messages using threshold cryptography and enqueue them in a local FIFO queue, ready to be sent out in the beginning of the next round. The \textit{when block} between lines 8-22 is executed by each miner in the beginning of a round. In line 9, miners arrive at consensus on the set of blocks proposed by each miner in the previous round. These blocks are stored in a set $candidate\_set$. Miners then clear $candidate\_set$, ready to store blocks in the current round and the consensus block $B$ is added to the blockchain. Lines 12-15 clear transactions contained in $B$ from the miners' memory pool, $MP$ (a set data structure containing transactions waiting to be added to the blockchain) and the causal past ($CP(t)$ keeps track of transactions in $MP$ that need to be added to the blockchain before $t$) of the remaining transactions in $MP$. Next, the miner constructs its own block (to be sent out for consensus in the next round) with semantically valid and safe transactions (lines 16-20). In lines 21-22, miners send their blocks for consensus in the next round. The \textit{when block} in lines 23-28 deals with incoming blocks from other miners for which consensus will be arrived at in the next round. When a miner receives a block, it checks if the block is semantically valid and makes sure all transactions in the block do not have causal dependencies on existing transactions in the memory pool. If that is the case, the block is added to $candidate\_set$. The \textit{when block} in lines 29-32 deals with miners receiving a message (containing a transaction) from a client via {\sf BR\_delivery}, computing their decryption shares for the message and broadcasting the decryption share in the next round. Finally, the \textit{when block} in lines 33-38 deals with miners receiving the required number of decryption shares ($t+1$) for decrypting a protocol message. The miners decrypt the message $m$ in line 36 and extract the transaction $t_m$ in line 37 (this line is {\sf BC\_delivery}) and store the causal past of $t_m$ in $CP(t_m)$. Finally, $t_m$ is added to the memory pool in line 39. For the purposes of this protocol, $CP(t)$ is treated as a dynamic set data structure, which starts off containing the entire set of transactions in the causal past of $t$. As each of these transactions are added to blockchain $BT$, they are removed from $CP(t)$. Once $CP(t) = \phi$, $t$ is a safe transaction and is ready to be added to the blockchain. 

\begin{lemma}

\label{lm:cp}

Memory pool, $MP$, will be the same at all correct miners following Algorithm \ref{alg:co_protocol} at the end of every round.
    
\end{lemma}

\begin{proof}

All $(2t+1)$ correct miners' decryption shares are required to decrypt any $C_m$ (ciphertext of message $m$) that has been {\sf BR\_deliver}ed as seen in line 33. Therefore, no miner can see transaction $t_m$ (contained in $m$) until the last miner to {\sf BR\_deliver} $m$ sends its decryption share in lines 29-32. If the last miner to broadcast its decryption share for $C_m$ does so in round $r$, all correct miners will {\sf BC\_deliver} $m$ and add $t_m$ to their mining pools in round $r$ (lines 32-38). Therefore, at the end of round $r$, $MP$ will be the same at all correct miners. $\qed$
    
\end{proof}

\begin{theorem}
    \label{th:algo_co_safety}
     For all transactions $t_1$ and $t_2$ in a valid blockchain $BT$, such that $t_1 \rightarrow t_2$, Algorithm \ref{alg:co_protocol} guarantees that $t_1$ is ordered before $t_2$ in $BT$'s consensus chain.

\end{theorem}

\begin{proof}

Consider messages $m_1$ and $m_2$ containing transactions $t_1$ and $t_2$ respectively, with $m_1 \rightarrow m_2$. From Definition \ref{def:hb_tx}, $t_1 \rightarrow t_2$. Let $p_{m_j}$ (possibly Byzantine) be the sender of $m_2$. $p_{m_j}$ {\sf BC\_deliver}s $m_1$ and views $t_1$ in line 37 of Algorithm \ref{alg:co_protocol}. The earliest that $m_2$ can be broadcasted to the system is in round $r$ itself (this is Byzantine behaviour, a correct miner would broadcast $m_2$ in round $(r+1)$). The fastest delivery time of $m_2$ at any miner would be the minimum latency of BRB ($\gamma$) + decryption share latency (1 round) + sending round ($r$). Therefore, the earliest $m_2$ can be delivered at any miner is at round $r_{m_2} = (r + \gamma + 1)$. The latest delivery of $m_1$ at any miner would be round $r_{m_1} = (r + \beta - \gamma)$. For any BRB protocol, $\beta < 2\gamma$, therefore we can derive the following by replacing $\beta$ with $2\gamma$ in the equation for $r_{m_1}$: 

\begin{center}

 $r_{m_1} < (r + \gamma)$ \\
 $r_{m_1} < (r + \gamma + 1)$ \\
 $r_{m_1} < r_{m_2}$
    
\end{center}

Since $r_{m_1} < r_{m_2}$, $m_1$ will be {\sf BC\_deliver}ed before $m_2$ and $t_1$ will be in the memory pool ($MP$) prior to the extraction of $t_2$ (lines 33-38). Therefore, when $t_2$ will be included in $MP$ at all correct miners, $CP(t_2)$ will include $t_1$ (lines 36-38). From Lemma \ref{lm:cp}, $t_1 \in CP(t_2)$ at all correct miners (line 37). Any block $B$ containing $t_2$ will be rejected by correct miners if $t_1$ is not recorded in blockchain $BT$ (lines 23-26). Consequently, no such block $B$ can be added to the blockchain until $t_1$ is recorded in $BT$. Therefore, given $t_1 \rightarrow t_2$, $t_1$ will be recorded in $BT$ prior to $t_2$.$\qed$

\end{proof}

\begin{theorem}
    \label{th:algo_co_liveness}
     All transactions broadcasted to blockchain $BT$ via Algorithm \ref{alg:co_protocol} will be added to each correct miner's memory pool $MP$ within bounded time.
\end{theorem}

\begin{proof}

Let client $p_{c_i}$ send message $m$ (containing transaction $t_m$) to $BT$ via Algorithm \ref{alg:co_protocol} in round $r$. $p_{c_i}$ sends $m$'s ciphertext $C_m$ via BRB in lines 1-4 to the system of miners. By BRB-Termination-1 and BRB-Termination-2 from Definition \ref{definition:BRB0}, it can be seen that all correct miners will {\sf BR\_deliver} $C_m$ within $\beta$ rounds at line 29 and broadcast their respective decryption shares in the next round in lines 30-32. In the next round all correct processes will receive the required number of decryption shares to decrypt $C_m$ in line 33. All correct miners will proceed to decrypt message $m$ and store its transaction $t_m$ in $MP$ in the same round (lines 34-38). Therefore, a transaction $t_m$ sent in round $r$ via Algorithm \ref{alg:co_protocol} will arrive at every correct miner's memory pool during or before round $(r+\beta+1)$. $\qed$

\end{proof}

\begin{corollary}
    \label{co:causal}

    Algorithm \ref{alg:co_protocol} guarantees causal ordering as defined in Definition \ref{definition:ss+l_tx}.
    
\end{corollary}

\begin{theorem}
    \label{th:fr_sec}

     Any blockchain constructed by Algorithm \ref{alg:co_protocol} is resilient to front-running attacks.
    
\end{theorem}

\begin{proof}
    Follows from Theorem \ref{th:front_run} and Corollary \ref{co:causal}.
\end{proof}

A critical observation about Algorithm \ref{alg:co_protocol} is that any transaction $t$ {\sf BC\_deliver}ed in round $r$ will be added to the causal past of every transaction {\sf BC\_deliver}ed in rounds $(r+1),(r+2),...(r+k)$, where $(r+k)$ is the round where $t$ is recorded to the blockchain. Consequently, any transaction $t'$ added to the memory pool after $t$ cannot be added to the blockchain until $t$ is added to it. This forces miners to mine and add existing transactions in the memory pool to the blockchain in order to ensure that future transactions do not end up waiting in the memory pool, thereby preventing wastage of both resources and time. This leads us to Observation \ref{ob:fairness}.

\begin{observation}
    \label{ob:fairness}

    Any blockchain constructed by Algorithm \ref{alg:co_protocol} guarantees intrinsic fairness to clients.

\end{observation}

\end{section}

\begin{section}{Discussion}

\textbf{Front-running attacks}. In this paper we studied front-running attacks and proved that all front-running attacks are causal ordering violations accross transactions.
The reason that front-running attacks are feasible against existing block\-chains is because blockchains provide a \textit{total ordering} of transactions by solving Byzantine-tolerant consensus but do not preserve causality when building this total-ordering. We conclude that solving consensus is not enough from an application semantics perspective in a Byzantine environment. 

\textbf{Stronger Blockchains}. In light of our findings, we defined the notion of a \textit{strong blockchain}, which is a blockchain that provides a causality-preserving total-order across transactions. This eliminates the feasibility of front-running attacks by Byzantine processes and guarantees application semantics. We proposed a causal ordering protocol to be used in conjunction with the consensus protocol to build a strong blockchain. This approach is modular because it does not interfere with the consensus protocol of the blockchain. Instead, the causal ordering protocol on transactions runs prior to the consensus protocol on blocks of transactions. That is, the causal ordering protocol ensures that transactions added to blocks do not have causal dependencies in the memory pool. This makes it straightforward to incorporate causal ordering as a pre-consensus protocol to existing blockchains. Our blockchain protocol keeps track of causal dependencies of every transaction added to the memory pool of every miner. BRB ensures that all correct miners have correct knowledge of the causal dependencies. This allows our protocol to stop any transactions from being mined whose causal dependencies have not been added to the blockchain. It only allows blocks containing \textit{safe transactions} to be candidates for consensus.

\textbf{Related Work.} Recently, a technique to make sandwich attacks unprofitable to \textit{rational} Byzantine processes in the permissionless setting was proposed \cite{alpos2023eating}. This technique involves changing the blockchain protocol itself by making random reorderings of transactions within proposed blocks. Fair ordering of transactions at the consensus level has been formalized in \cite{kelkar2022order,cachin2022quick}. However, this approach does not completely rule out front-running attacks.
Our protocol uses threshold cryptography, which has previously been used in a probabilistic algorithm based on atomic (total order) broadcast for secure causal atomic broadcast (liveness and strong safety) in an asynchronous system \cite{cachin2001secure}. This algorithm used acknowledgements and effectively processed the atomic broadcasts serially. This protocol would force miners to see transactions in a total order inhibiting parallel mining of transactions sent concurrently. Additionally, this protocol in conjunction with blockchain would solve consensus twice, wasting time and resources. More recently, threshold cryptography has been used to develop a non-deterministic multicast algorithm for causal ordering in asynchronous systems \cite{misra2023byzantine2}.

\textbf{Causality Preserving Blockchain Protocol}. We proposed a strong block\-chain protocol and proved its correctness in this paper. Our protocol provides \textit{deterministic} causal ordering in a synchronous communication model. Since our protocol operates in a synchronous setting, the consensus protocol will also be deterministic. Our protocol assumes that there are $(3t + 1)$ miners out of which at most $(t-1)$ can be Byzantine \footnote{BRB requires an upper bound of $t$ Byzantine processes out of ($3t+1$) processes. In our case, the client becomes the $(3t+2)^{th}$ process in the system when broadcasting to the system of miners via BRB. In case the broadcasting client is Byzantine, correctness of the protocol can only be guaranteed when at most $(t-1)$ miners are Byzantine.}. This means that this protocol is suited for a permissioned blockchain, with a static number of miners. Our protocol has a message complexity of $O(n^2)$ and has an upper bound on latency (time for a transaction to arrive in all correct miners' memory pools) of $(\gamma + 1)$ rounds.

\textbf{Conclusion}. Our result in Theorem \ref{th:front_run}, stating that front-running attacks are causal violations is independent of the system model of our protocol. Therefore, front-running attacks are not feasible against our notion of a strong blockchain regardless of the system model assumptions (permissioned vs. non-permissioned, synchrony vs. asynchrony). Future work comprises developing protocols for strong blockchain in different system settings such as non-permissioned blockchains and blockchains with asynchronous communication. Although our solution is deterministic, it is not possible to develop a deterministic strong blockchain in an asynchronous system \cite{misra2023byzantine}. This paper established that causal ordering is critical for blockchain security and maintaining application semantics and provided a causal ordering solution for synchronous permissioned blockchains. To the best of our knowledge, this is the first work that addressed the \textit{root cause} that makes front-running attacks possible, and proved that front-running attacks are causal violations. Additionally, we provided a solution that can be adopted by existing blockchains \textit{without interfering} with the blockchain protocol.

\end{section}

\bibliographystyle{splncs04}
\bibliography{references}

\begin{thebibliography}{10}
\providecommand{\url}[1]{\texttt{#1}}
\providecommand{\urlprefix}{URL }
\providecommand{\doi}[1]{https://doi.org/#1}

\bibitem{alpos2023eating}
Alpos, O., Amores-Sesar, I., Cachin, C., Yeo, M.: Eating sandwiches: Modular and lightweight elimination of transaction reordering attacks. arXiv preprint arXiv:2307.02954  (2023)

\bibitem{bracha1987asynchronous}
Bracha, G.: Asynchronous byzantine agreement protocols. Information and Computation  \textbf{75}(2),  130--143 (1987)

\bibitem{bracha1985asynchronous}
Bracha, G., Toueg, S.: Asynchronous consensus and broadcast protocols. Journal of the ACM (JACM)  \textbf{32}(4),  824--840 (1985)

\bibitem{cachin2001secure}
Cachin, C., Kursawe, K., Petzold, F., Shoup, V.: Secure and efficient asynchronous broadcast protocols. In: Annual International Cryptology Conference. pp. 524--541. Springer (2001)

\bibitem{cachin2022quick}
Cachin, C., Mi{\'c}i{\'c}, J., Steinhauer, N., Zanolini, L.: Quick order fairness. In: International Conference on Financial Cryptography and Data Security. pp. 316--333. Springer (2022)

\bibitem{dickerson2017adding}
Dickerson, T., Gazzillo, P., Herlihy, M., Koskinen, E.: Adding concurrency to smart contracts. In: Proceedings of the ACM Symposium on Principles of Distributed Computing. pp. 303--312 (2017)

\bibitem{DBLP:journals/jacm/DworkLS88}
Dwork, C., Lynch, N.A., Stockmeyer, L.J.: Consensus in the presence of partial synchrony. J. {ACM}  \textbf{35}(2),  288--323 (1988)

\bibitem{eskandari2020sok}
Eskandari, S., Moosavi, S., Clark, J.: Sok: Transparent dishonesty: front-running attacks on blockchain. In: Financial Cryptography and Data Security: FC 2019 International Workshops, VOTING and WTSC, St. Kitts, St. Kitts and Nevis, February 18--22, 2019, Revised Selected Papers 23. pp. 170--189. Springer (2020)

\bibitem{hahn2017smart}
Hahn, A., Singh, R., Liu, C.C., Chen, S.: Smart contract-based campus demonstration of decentralized transactive energy auctions. In: 2017 IEEE Power \& energy society innovative smart grid technologies conference (ISGT). pp.~1--5. IEEE (2017)

\bibitem{hanifatunnisa2017blockchain}
Hanifatunnisa, R., Rahardjo, B.: Blockchain based e-voting recording system design. In: 2017 11th International Conference on Telecommunication Systems Services and Applications (TSSA). pp.~1--6. IEEE (2017)

\bibitem{imbs2016trading}
Imbs, D., Raynal, M.: Trading off t-resilience for efficiency in asynchronous byzantine reliable broadcast. Parallel Processing Letters  \textbf{26}(04),  1650017 (2016)

\bibitem{kelkar2022order}
Kelkar, M., Deb, S., Kannan, S.: Order-fair consensus in the permissionless setting. In: Proceedings of the 9th ACM on ASIA Public-Key Cryptography Workshop. pp. 3--14 (2022)

\bibitem{lamport2019byzantine}
Lamport, L., Shostak, R., Pease, M.: The byzantine generals problem. In: Concurrency: the works of leslie lamport, pp. 203--226 (2019)

\bibitem{DBLP:journals/toplas/LamportSP82}
Lamport, L., Shostak, R.E., Pease, M.C.: The byzantine generals problem. {ACM} Trans. Program. Lang. Syst.  \textbf{4}(3),  382--401 (1982)

\bibitem{misra2022solvability}
Misra, A., Kshemkalyani, A.D.: Solvability of byzantine fault-tolerant causal ordering problems. In: International Conference on Networked Systems. pp. 87--103. Springer (2022)

\bibitem{misra2023byzantine2}
Misra, A., Kshemkalyani, A.D.: Byzantine fault-tolerant causal order satisfying strong safety. In: International Symposium on Stabilizing, Safety, and Security of Distributed Systems. pp. 111--125. Springer (2023)

\bibitem{misra2023byzantine}
Misra, A., Kshemkalyani, A.D.: Byzantine fault-tolerant causal ordering. In: Proceedings of the 24th International Conference on Distributed Computing and Networking. pp. 100--109 (2023)

\bibitem{nakamoto2008bitcoin}
Nakamoto, S.: Bitcoin: A peer-to-peer electronic cash system. Decentralized business review  (2008)

\bibitem{DBLP:journals/jacm/PeaseSL80}
Pease, M.C., Shostak, R.E., Lamport, L.: Reaching agreement in the presence of faults. J. {ACM}  \textbf{27}(2),  228--234 (1980)

\bibitem{polge2021permissioned}
Polge, J., Robert, J., Le~Traon, Y.: Permissioned blockchain frameworks in the industry: A comparison. Ict Express  \textbf{7}(2),  229--233 (2021)

\bibitem{saad2019exploring}
Saad, M., Spaulding, J., Njilla, L., Kamhoua, C., Shetty, S., Nyang, D., Mohaisen, A.: Exploring the attack surface of blockchain: A systematic overview. arXiv preprint arXiv:1904.03487  (2019)

\bibitem{shoup2002securing}
Shoup, V., Gennaro, R.: Securing threshold cryptosystems against chosen ciphertext attack. Journal of Cryptology  \textbf{15}(2),  75--96 (2002)

\bibitem{torres2021frontrunner}
Torres, C.F., Camino, R., et~al.: Frontrunner jones and the raiders of the dark forest: An empirical study of frontrunning on the ethereum blockchain. In: 30th USENIX Security Symposium (USENIX Security 21). pp. 1343--1359 (2021)

\bibitem{wood2014ethereum}
Wood, G., et~al.: Ethereum: A secure decentralised generalised transaction ledger. Ethereum project yellow paper  \textbf{151}(2014),  1--32 (2014)

\bibitem{zust2021analyzing}
Z{\"u}st, P., Nadahalli, T., Wattenhofer, Y.W.R.: Analyzing and preventing sandwich attacks in ethereum. ETH Z{\"u}rich  (2021)

\end{thebibliography}

\end{document}